\documentclass{article}
\usepackage[a4paper]{geometry}
\usepackage{amsmath,amssymb,graphicx,graphics}
\begin{document}

\begin{center}
\textbf{\large Influence of Elastic Strains on The Adsorption Process in Porous Materials. An Experimental Approach.}

Annie Grosman* and Camille Ortega

\textit{Institut des NanoSciences de Paris (INSP), Universit\'e Paris 6, UMR-CNRS 75-88, Campus Boucicaut, 140 rue de Lourmel, 75015 Paris, France}

\begin{abstract}
\noindent
The experimental results presented in this paper show the influence of the elastic deformation of porous solids on the adsorption process. With p$^{+}$-type porous silicon formed on highly boron doped (100) Si single crystal, we can make identical porous layers, either supported by or detached from the substrate. The pores are perpendicular to the substrate. The adsorption isotherms corresponding to these two layers are distinct. In the region preceding capillary condensation, the adsorbed amount is lower for the membrane than for the supported layer and the hysteresis loop is observed at higher pressure. We attribute this phenomenon to different elastic strains undergone by the two layers during the adsorption process. For the supported layer, the planes perpendicular to the substrate are constrained to have the same interatomic spacing as that of the substrate so that the elastic deformation is unilateral, at an atomic scale, and along the pore axis. When the substrate is removed, tridimensional deformations occur and the porous system can find a new configuration for the solid atoms which decreases the free energy of the system adsorbate-solid. This results in a decrease of the adsorbed amount and in an increase of the condensation pressure. The isotherms for the supported porous layers shift toward that of the membrane when the layer thickness is increased from $30$ to $100$~$\mu$m. This is due to the relaxation of the stress exerted by the substrate as a result of the breaking of Si-Si bonds at the interface between the substrate and the porous layer. The membrane is the relaxed state of the supported layer.
\end{abstract}
\end{center}

Author E-mail address : annie.grosman@insp.jussieu.fr

\section{Introduction}
\paragraph{}
The elastic deformation of porous materials during the gas adsorption-desorption process has been studied for a long time.\cite{Sereda67} The reversible formation of an adsorbed film preceding capillary condensation leads to an extension of the porous material, while, when condensation occurs, a contraction is observed, attributed to the negative fluid pressure under the concave menisci. As the vapor pressure is further increased up to the saturation vapor pressure, an extension occurs due to the vanishing of the negative liquid pressure consequent upon the flattening of the concave menisci. During the evaporation process, the hysteresis phenomenon is observed in the same pressure region as the adsorption phenomenon. \cite{Wiig49, Amberg52, Dolino96, Scherer01, Herman06} Similar results have been obtained in SBA-15 \cite{Zickler} and in MCM-41 \cite{Gunther} by means of in situ small-angle X-ray diffraction experiments.

Curiously, these deformations were generally ignored in the thermodynamics of adsorption. How does the elastic deformation of a porous solid affect the adsorption and capillary condensation? In the recent paper quoted above, \cite{Gunther} the authors address the question by means of Monte Carlo simulations. They found that the elastic deformation causes a shift of the order of a few percent of the phase transition toward that in the bulk. 

Simultaneously with this work, \cite{Gunther} using a classical thermodynamic approach, we have shown \cite{Grosman08-II} that the assumption of an inert adsorbent in adsorption is totally unjustified. In the physics of adsorption, the equilibrium condition for mass transfer is not given by the equality of the chemical potentials of the vapor and adsorbate but includes also the deformation of the solid. This allows us to introduce in the thermodynamic relationships, in addition to the classical surface free energy, the elastic energy stored in the solid, which is, for small deformations, quadratic in the deviation. The mechanical equilibrium obtained by the minimization of the free energy of the adsorbate-solid system with respect to the deformation results in the linear variation of the deformation with the surface free energy observed by some investigators, \cite{Amberg52, Bangham30, Dash76} with the proportionality factor depending on the elastic constants of the solid.

The linear variation of the deformation with the surface free energy is hence the result of two equilibrium conditions, with one being thermodynamic and the other mechanic. This important result means that the adsorbed amount in a given pore can be changed by a mechanical stress external to the pore.

This is the basis of the interaction mechanism we have proposed. \cite{Grosman08-II} The filling or the draining of a pore causes elastic deformation not only on its inner pore walls but also on the inner walls of its neighbors and thus causes a change of their surface free energy, i.e., a change of the adsorbed amount in a neighbor pore not yet filled and a change of the negative liquid pressure through the shape of the meniscus in a neighbor pore already filled. This interaction mechanism acts as an alternative to the pore blocking/percolation mechanism, \cite{Mason83, Mason88} which fails to explain the interpendence between pores observed in systems where there is no pore-pore intersection. \cite{Grosman05, Grosman08-I}

In the present paper, we find the same idea as that exposed in our interaction mechanism except that here the external mechanical stress is not imposed on a given pore by the neighbors but imposed by the substrate to which the porous layer is attached, which is the case for porous silicon. We thus come back to a problem which we have already addressed in a letter: \cite{CoasnePRL} the comparison of the adsorption isotherms for p$^{+}$-type porous silicon layers supported by the substrate with pores closed at one end and detached from the substrate (a membrane) with pores open at both ends. Capillary condensation systematically occurs at higher pressure in the membrane than in the supported layer. Following the idea of Cohan, \cite{Cohan38} we attributed this phenomenon to two different scenarios for the filling of these two porous systems. In a pore closed at one end, a preferential condensation occurs at the bottom of the pore, while in a pore open at both ends the adsorbed film becomes unstable and the filling occurs at higher pressure when a liquid bridge is formed somewhere in the pore.

In the light of our thermodynamic approach \cite{Grosman08-II} and of new experimental results presented in this paper, we reconsider this explanation. After discussing the similarity of the supported layer and the corresponding membrane in Section~\ref{subsec:layerandmemb}, we show in Section~\ref{subsec:elasticdef} that the difference between the two isotherms is due to the different elastic deformations undergone by the two layers on account of the stress exerted by the substrate: for the supported layer, the deformation is unilateral (perpendicular to the substrate) while the membrane is submitted to tridimensional deformations not necessarily isotropic. To these different configurations for the solid atoms correspond different free energies. In Section~\ref{subsec:interfstress}, we show that this effect disappears gradually when increasing the layer thickness from $30$ to $100$~$\mu$m: the isotherm corresponding to the supported layer shifts toward that of the membrane. This is due to the relaxation of the stress exerted by the substrate as some Si-Si bonds are broken at the porous layer-substrate interface by the stresses generated at this interface. In Section~\ref{sec:discussion} we will discuss these whole results.
\section{Experimental Section}
\label{sec:exp}
\subsection{Preparation and Properties of p$^{+}$-type Porous Silicon}
\paragraph{}
The preparation process, the morphology, and the physical properties of p$^{+}$-type porous silicon have been reported in detail in recent papers. \cite {CoasnePRL, Grosman08-I} We give here a brief outline of these properties. p$^{+}$-Type porous Si is obtained by anodic dissolution, at constant current, of highly boron doped (100) Si single crystal in hydrofluoric acid (HF) solution. It is a porous single crystal material. The pores are perpendicular to the Si substrate, and all have the same length. Transmission electron microscopy (TEM) measurements show that (i)- at the surface of the porous layer, the pore cross sections are of polygonal shape, (ii)- the pore size distribution (PSD), which depends on the formation conditions (current density and HF concentration), is large ($\sim13\pm6$~nm and $\sim26\pm14$~nm for a $50\%$ and $70\%$ porosity layer, respectively), and (iii)- the thickness of the Si walls separating the pores is almost constant. The two-dimensional (2D) TEM images yield the same porosity as that obtained by weighing. Moreover, the pore volume measured by weighing or by adsorption measurement once the pores are full of liquid is proportional to the etching time. This indicates that during the formation process the Si walls become insulating once a certain thickness is reached, $5-6$~nm in our case, even though the dopant atoms are not removed from the porous layer during the formation process. \cite{Grosman97} Thus, except for the presence of numerous facets on the Si walls, \cite{Bardeleben93} i.e., a roughness at the scale of few atoms, the pore cross-sectional areas are constant in depth. The cause of this high resistivity is not yet well understood. What we know is that the Si-HF interface plays a fundamental role in this property, since the Si walls remain electrically resistive as long as the HF environment is present. To render the Si walls electrically conducting, we must replace the acid HF solution by a neutral solution; thus, anodic oxidation occurs in place of anodic dissolution. This electrical property can be used to make membranes. At the end of the preparation process, once the desired pore length is obtained, the anodic current can be either cut off to form a porous layer supported by the Si substrate with pores closed at one end or increased to go into an electropolishing regime during which the Si walls are dissolved at the bottom of the pores to form a membrane with pores open at both ends. For this paper, we have prepared porous layers with a porosity of $50\%$ and of different thicknesses varying between $10$ and $100$~$\mu$m. We have also prepared the corresponding membranes. The electrochemical conditions were HF($40\%$)/EtOH=3:1, $J=20$~mA/cm$^{2}$, and etching rate $\sim1$~$\mu$m/min. The main structural parameters of this kind of layer can be found in ref~15.
\subsection{Measurements}
\paragraph{}
The nitrogen adsorption isotherm measurements were made at $77.4$~K using a Micromeritics ASAP2010 instrument equipped with pressure transducers with full scales of $1$~$\mu$mHg, $10$~mmHg, and $1000$~mmHg.

The radius of curvature of the supported porous layers was determined optically by measuring the diameters of Newton's rings. For these measurements, we have prepared porous layers on Si wafers polished on both sides. The interference patterns were caused by the difference in path lengths of the reflecting light from the concave back side of the Si substrate and from a flat mirror, with the whole device being in the configuration of a Michelson interferometer with equal arms. We used a red laser light (wavelength of 0.6328~$\mu$m).

The cross section view of the porous layer-silicon substrate interface was obtained using the high spatial resolution of a Field Emission Gun Scanning Electron Microscope (FEG-SEM). The beam energy was $5$~kV and the magnification was $200000\times$.
\section{Results and Data Analysis}
\label{sec:result}
\subsection{The porous layers supported by the substrate and the corresponding membranes}
\label{subsec:layerandmemb}
\begin{figure}
\begin{center}
\includegraphics[width=8.5 cm]{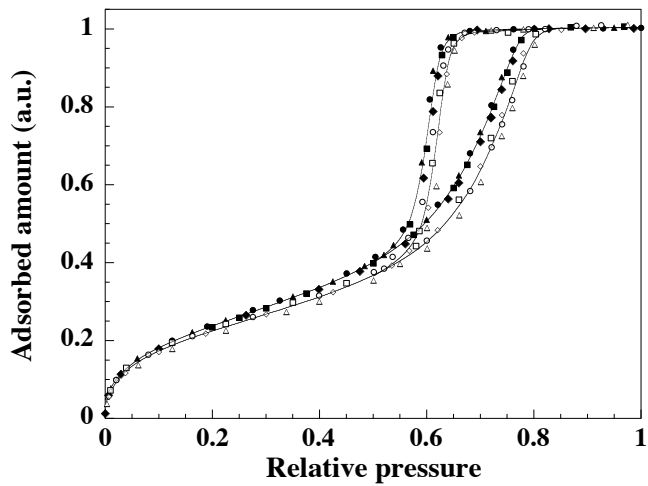}
\end{center}
\caption{Nitrogen adsorption isotherms at $77.4$~K for several p$^{+}$-type porous silicon layers supported by the Si substrate (symbols: $\blacksquare$, $\bullet$, $\blacktriangle$, $\blacklozenge$) and the corresponding membranes (symbols: $\square$, $\circ$, $\triangle$, $\lozenge$). The solid lines are interpolations of the mean values. The thickness equals $20$~$\mu$m, and the porosity is $50\%$.}
\label{fig:layermemb}
\end{figure}

\paragraph{}
As noted in the Introduction, we come back, in this paper, to a problem which we have already addressed: the comparison of the adsorption isotherms for porous layers supported by the substrate with pores closed at one end and for self-supported layers with pores open at both ends. Figure~\ref{fig:layermemb} shows nitrogen adsorption isotherms at $77.4$~K, normalized to the same pore volume, for several $20$~$\mu$m thick supported layers and the corresponding membranes, with porosity of $50\%$. The solid lines are interpolations of the mean values of the two types of layers. Note that when making membranes, it is difficult to collect all the sample pieces, with some of them being too small and others, located at the periphery of the layer, remaining attached to the substrate. It is the reason why the isotherms for the supported layer and the corresponding membrane were normalized to the same pore volume.

The condensation branch was systematically observed at higher pressure for the membrane than for the supported porous layer. Following the idea of Cohan, \cite{Cohan38} we attributed \cite{CoasnePRL} this phenomenon to the effect of pore ends on the condensation process.

However, two results shown in Figure~\ref{fig:layermemb} remained unexplained. (1) In the region of reversible adsorption preceding the capillary condensation ($p\leqslant 0.5$), the adsorbed amount is systematically lower for the membrane than for the supported layer. (2) The evaporation branches of the layer and the corresponding membrane are not superimposed.

Under the usual assumption of an inert adsorbent, the presence of a second opening in otherwise identical porous layers, where the pore cross section is invariant with depth, should change neither the adsorbed amount before condensation nor the position of the evaporation branch on the pressure axis.

As the results of Figure~\ref{fig:layermemb} are the central point of the paper, we find it important to come back to the problem of the strict similarity of the supported and detached layers.

The only difference between these two layers, if any, could be a shift of the whole PSD equal to a fraction of the pore wall thickness, for example, $\sim1$~nm, under the effect of the electropolishing process. This process, which dissolves the Si walls at the bottom of the pores, could slightly dissolve the Si walls in the whole porous system, increasing the pore size. The determination of the PSD using TEM, described in detail in refs~15 and 16, is not sufficiently precise to measure such a shift. Our PSD measurements concern the supported layers.

We have shown \cite {Grosman08-I} that the mean pore diameters defined by $<D_{p}> =<p>/\pi$, where $<p>$ is the mean pore perimeter, and by $<D_{s}>=(4s/\pi)^{1/2}$, where $<s>$ is the mean surface area of the pore section, are quite similar ($<D_{p}>=13.2$~nm and $<D_{s}>=13.3$~nm). Thus, the $S/V$ ratio of the pore surface area $S=NL<p>$ over the pore volume $V=NL <s>$ is inversely proportional to the common value $<D>$ ($N$ is the number of pores, and $L$ is the pore length). The adsorbed amount before capillary condensation normalized to the same pore volume is hence inversely proportional to $<D>$, so that, if we suppose that the deformation of the adsorbent does not play any role on the adsorption process, the comparison of the normalized adsorbed amounts before capillary condensation for the supported layer and membrane is an excellent criterion to decide whether these two layers are identical. The lower is the normalized adsorbed amount before capillary condensation, the higher is the mean pore size.

Thus, under the assumption of an inert adsorbent, the conclusion of the results of Figure~\ref{fig:layermemb} is that the mean pore size of the $20$~$\mu$m thick membrane is higher than that of the corresponding supported layer. Concerning the condensation process, the shift of the condensation branch toward higher pressures observed for the membrane could be explained either by the sole increase of the pore size without invoking any influence of pore end on the condensation process or by the addition of the two effects.
\begin{figure}
\begin{center}
\includegraphics[width=8.5 cm]{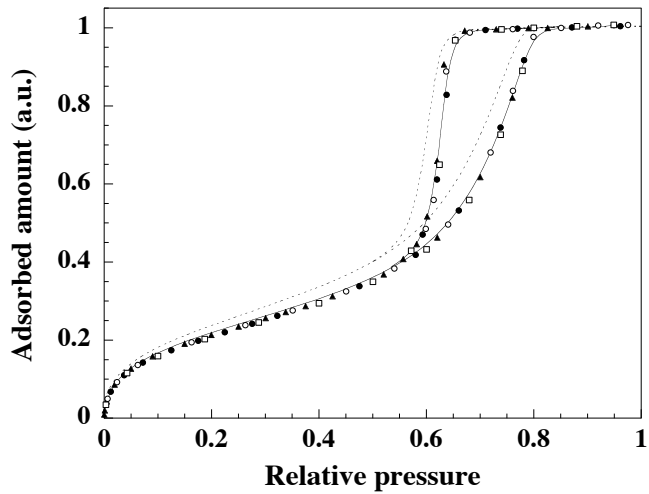}
\end{center}
\caption{Nitrogen adsorption isotherms at $77.4$~K for $100$~$\mu$m thick supported porous silicon layers (symbols:$\bullet$, $\blacktriangle$) and the corresponding membranes (symbols:$\circ$, $\square$). The dashed and solid lines are, respectively, the interpolations of the mean values for the $20$~$\mu$m thick supported porous layers and membranes represented in Figure~\ref{fig:layermemb}. The porosity is $50\%$.}
\label{fig:isot100}
\end{figure}

Nevertheless, the above explanation seemed to us very unlikely. The electrochemical etching rate is typically a few $\mu$m/min and the Si walls are $5-6$~nm thick. The simple fact that we can obtain membranes of different thicknesses by this process, with reproducible normalized adsorption-desorption isotherms (see below), without paying special attention to the value of the current intensity chosen in the large electropolishing regime, $\geqslant 100$~mA/cm$^{2}$ in our experimental conditions, strongly suggests that the porous layer remains unchanged during the electropolishing process.

Fortunately new results made it possible to solve this problem. Indeed, when making thicker supported layers ($100$~$\mu$m) and the corresponding membranes, we fortuitously fell upon a surprising result. As shown in Figure~\ref{fig:isot100}, the adsorption-desorption isotherm corresponding to a $100$~$\mu$m thick supported layer is fully identical to that of the corresponding membrane, and in addition to that of the $20$~$\mu$m thick membrane.
The experiment was repeated several times with similar results.

Two conclusions can be drawn. 

(i) The isotherms corresponding to the thin and thick membranes are superimposed once normalized to the same pore volume, which shows that they differ only by their thickness. If the electropolishing process used to obtain the two membranes modifies the PSDs, the modification is independent of the porous layer thickness. In the region preceding capillary condensation, the adsorption isotherms corresponding to the $100$~$\mu$m thick supported layers are superimposed on those corresponding to the two membranes. Making the same assumption of an inert adsorbent as for Figure~\ref{fig:layermemb}, we conclude that the electropolishing process lets the $100$~$\mu$m thick supported layers unchanged. This is hence also the case for the $20$~$\mu$m thick supported layer so that it is similar to its corresponding membrane. This is in contradiction with the above interpretation of Figure~\ref{fig:layermemb}.

(ii) The isotherms are also superimposed in the region of capillary condensation, which means that the thick supported layers behave like the membranes, which is not the case for the thin supported layers. Concerning this point, the question is whether this could be simply explained by the disappearance of pore end effect on the condensation process when making thicker supported layers. If we look at the change of the adsorption isotherm when increasing the thickness of the supported layer from $20$ to $100$~$\mu$m we note that the shift of the condensation branch toward higher pressures is accompanied by a decrease of the normalized adsorbed amount before capillary condensation. This indicates that this shift is not only due to pore end effect, if any. Indeed, still under the assumption of an inert adsorbent, pore end effect can influence the condensation process but should not affect the adsorbed amount before condensation.

We conclude that the assumption of an inert adsorbent during the adsorption process must be abandoned if we want to explain the whole results shown in Figures~\ref{fig:layermemb} and \ref{fig:isot100}. So, in the following section, we will analyze the above striking results by taking into account the elastic deformation of the porous matrix. At this stage, since the adsorbent cannot be regarded as inert anymore, the comparison of the different isotherms does not allow us to draw a conclusion concerning the similarity of the supported layers and the corresponding membranes. In what follows, we have chosen to assume their strict similarity, a hypothesis which seemed to us by far the most plausible and which will be validated a posteriori.

In Section~\ref{subsec:elasticdef}, we first analyse the results shown in Figure~\ref{fig:layermemb}, and, in Section~\ref{subsec:interfstress}, those of Figure~\ref{fig:isot100}. The possible effect of pore ends on the condensation process will be discussed. 
\subsection{Elastic deformation of the porous layers}
\label{subsec:elasticdef}
\paragraph{}
The lattice parameter of porous silicon layers is different from that of silicon: \cite{Barla} porosity affects the crystal lattice by producing a slight expansion. For p$^{+}$-type porous layers similar to those we study (highly boron doped (100) Si wafer, porosity $\simeq50\%$), the lattice parameter of porous layers supported by the substrate (subscript $sl$), measured in the [100] direction, is bigger than that of the substrate: $(\Delta a/a)_{sl}=5.15\times10^{-4}$. For the corresponding membrane (subscript $m$) the lattice parameter (still in the [100] direction) is a little less large: $(\Delta a/a)_{m}=4.2\times10^{-4}$. This is due to the fact that, for the supported porous layers, the planes perpendicular to the substrate are constrained to have the same interatomic spacing as that of the substrate, leading to an additional expansion: for the layers studied by Barla et al., \cite{Barla} $(\Delta a/a)_{sl}=1.2~(\Delta a/a)_{m}$.

The expansion of porous silicon is attributed to the presence of SiH$_{x=1, 2, 3}$ groups chemically adsorbed during the formation process, \cite{Sugiyama, Dolino96-II} but, to our knowledge, nobody has measured the lattice parameter of clean porous layers. Once the layers are exposed to air, the hydrogen atoms are slowly replaced by oxygen over a timescale of a few months. \cite{Loni} The two $(\Delta a/a)$ values given above probably correspond to freshly prepared porous layers covered by the same amount of SiH$_{x}$ groups, i.e., to the same surface stress.

Thus, during adsorption experiments, \textit{for a given adsorbed amount}, we should have
\begin{equation}
\bigg(\frac{\delta a}{a}\bigg)_{sl}=\beta\bigg(\frac{\delta a}{a}\bigg)_{m},
\end{equation}
where $\delta a$ is the variation of the lattice parameter (in the [100] direction) caused by adsorption and $\beta>1$ is a constant which depends on the morphology of the porous layer. $\beta=1.2$ for the layers of ref~20.

We have no information on the elastic deformation $(\delta a/a)$ of the supported porous layer and the corresponding membrane during N$_{2}$ adsorption experiment, but several investigators \cite{Amberg52, Bangham30, Dash76} have shown that, in the region preceding capillary condensation, the linear deformation of porous materials is related to the adsorbed amount by the equation 
\begin{equation}
\int_{0}^{P}N_{\sigma}\,k_{B}T\,\frac{dP}{P}=k\frac{\delta a}{a},
\end{equation}
where $N_{\sigma}$ is the number of the adsorbed molecules, $P$ is the vapor pressure, and $k$ is a proportionality factor which depends on the elastic constants of the porous solid.

We have shown \cite{Grosman08-II} that
\begin{equation}
\big|F_{\rm so}^{el}+\gamma A\big|_{0}^{P}=-\int_{0}^{P}N_{\sigma}\,k_{B}T\,\frac{dP}{P},
\end{equation}
where $F_{\rm so}^{el}$ is the elastic energy of the solid, $\gamma$ is the surface free energy per unit area, and $A$ is the surface area of the porous solid. The elastic energy stored in the solid during adsorption can be neglected compared to the variation of the surface free energy so that eq~3 becomes
\begin{equation}
\Delta(\gamma A)=\big|\gamma A\big|_{0}^{P}=-\int_{0}^{P}N_{\sigma}\,k_{B}T\,\frac{dP}{P}
\end{equation}
which is the classical Gibbs adsorption equation established under the assumption of an inert adsorbent.
From eqs~3 and 4, it follows 
\begin{equation}
\big|\gamma A\big|_{0}^{P}=\big[\gamma(a)-\gamma(a_{0})\big]A=-k\frac{\delta a}{a},
\end{equation}
where $a_{0}$ and $a$ are the lattice parameters before and after adsorption, respectively.

\begin{figure}
\begin{center}
\includegraphics[width=8.5 cm]{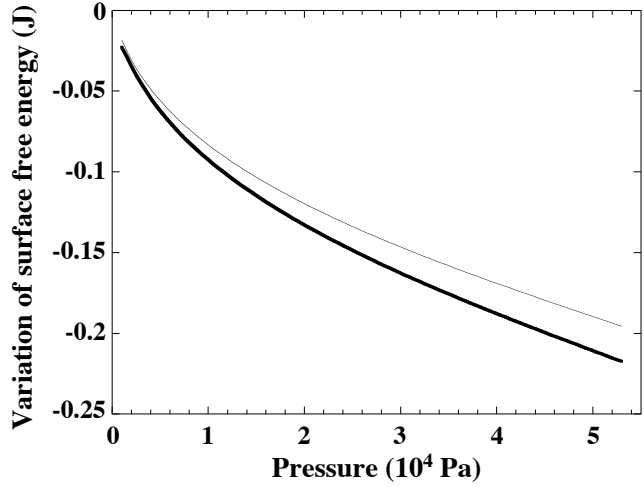}
\end{center}
\caption{Variation of the surface free energy as a function of the gas pressure, deduced from nitrogen adsorption isotherms at $77.4$~K in the region preceding the capillary condensation process, for a $20$~$\mu$m thick supported porous silicon layer (thick line) and its corresponding membrane (thin line). The mass of the porous layer equals $36.36\times10^{-3}$~g.}
\label{fig:surfreeE}
\end{figure}

The variations of the surface free energy as a function of the vapor pressure, corresponding to the two isotherms represented in Figure~\ref{fig:layermemb}, in the region preceding capillary condensation, are shown in Figure~\ref{fig:surfreeE}. For a given adsorbed amount, i.e., a given variation of the surface free energy, the elastic deformations of the supported layer and membrane are related by eq~1. It follows from eq~5 that
\begin{equation}
k_{m}=\beta~k_{sl},
\end{equation}
where $k_{m}$ and $k_{sl}$ are the proportionality factors, defined by eq~5, corresponding to the membrane and the supported layer, respectively.

The ratio $\Delta(\gamma _{sl}A)/\Delta(\gamma _{m}A)$ calculated from the data represented in Figure~\ref{fig:surfreeE} is independent of the vapor pressure and equals $1.111\pm0.001$.

Thus, \textit{at each vapor pressure} during the adsorption process, the extension of the supported layer and the membrane should obey the relation
\begin{equation}
\bigg(\frac{\delta a}{a}\bigg)_{sl}=1.11~\beta~\bigg(\frac{\delta a}{a}\bigg)_{m}.
\end{equation}
Taking $\beta=1.2$, for example, we obtain $(\delta a/a)_{sl}=1.33~(\delta a/a)_{m}$. A decrease of $\simeq30\%$ in ($\delta a/a$) along the pore axis accompanied by transversal deformations then leads to a significant difference ($10\%$) in the adsorbed amounts (see Figure~\ref{fig:layermemb}).

It remains now to verify whether the variation of the surface free energy for either porous layer corresponds to acceptable values of the relative deformation. It is easier to make this estimation for the supported layer, since the deformation is unilateral. We have shown \cite{Grosman08-II} that $k=CV_{\rm so}/\alpha$, where $C$ is defined by
\begin{equation}
F_{\rm so}^{el}= \frac{1}{2}\,C\,V_{\rm so}\,\bigg(\frac{\delta a}{a}\bigg)^{2}.
\end{equation}
$C$ is a constant which depends on Young's modulus $E_{p}$, on Poisson's coefficient $\nu_{p}$, and on the geometry of the porous solid. $V_{\rm so}$ is the porous solid volume, and $\alpha$ is a constant defined by $dA/A=\alpha (da/a)$.

Equation (5) becomes 
\begin{equation}
\big[\gamma(a)-\gamma(a_{0})\big]A=-\frac{CV_{\rm so}}{\alpha}~ \bigg(\frac{\delta a}{a}\bigg).
\end{equation}
For the supported porous layer, perpendicular to the pore axis $(\delta a/a)_{\bot}=0$. In this case, $dA/A=da/a$ and $\alpha=1$. According to the classical theory of elasticity, \cite{Landau} the constant $C$ is related to Young's modulus by the relation 
\begin{equation}
C=E_{p}~\frac{1-\nu_{p}}{1-\nu_{p}-2\nu_{p}^{2}}.
\end{equation}
As $\nu_{p}\simeq0.1$, \cite{Barla} $C\simeq E_{p}$ and $k_{sl}\simeq E_{p}~V_{\rm so}$.

Young's modulus values $E_{\rm p}$ of p$^{+}$-type porous silicon measured using different techniques \cite{Bellet-96-II} are close to $E$~$(1-P_{or})^{2}$, a relation found by Gibson and Ashby, \cite{Gibson} where $E=166$~GPa is Young's modulus of bulk silicon \cite{McSkimin} and $P_{or}=0.5$ is the porosity of our porous layers. Taking a density of $2.32$~g/cm$^{3}$ for Si, we find $k_{sl}=1.8\times10^{4}$~J/g. Considering the data represented in Figure~\ref{fig:surfreeE}, we have, at $P=5\times10^{4}$~Pa for instance, $\Delta(\gamma _{sl}A)=-0.21~J/36.36\times10^{-3}~g=-5.78$~J/g, which corresponds to $({\delta a/a})_{sl}=3.2\times10^{-4}$. 

It must be noted that the elastic deformations measured before or during adsorption are caused by the interaction adsorbate-porous solid and must be considered at an atomic scale. Hence, in adsorption experiments, Young's modulus may be higher than that given by the Gibson-Ashby relation which corresponds to the elastic response of a porous solid to an external loading, and the actual deformations could be lower than those estimated in this paper.

We now analyze the above results.

(1) First of all, let us compare the surface free energy to the other terms of the total free energy of the system adsorbate-porous solid which is given by \cite{Grosman08-II}: 
\begin{equation}
F_{\sigma,\rm so}= - P_{\sigma}V_{\sigma}+\mu_{\rm v}N_{\sigma}+F_{\rm so}^{el}+\gamma A.
\end{equation}
where $\gamma A\gg F_{\rm so}^{el}$ depends on the elastic deformation through the mechanical equilibrium eq~5. 
As long as we are concerned by the reversible adsorption of a few monolayers the term $P_{\sigma}V_{\sigma}$, where $P_{\sigma}$ and $V_{\sigma}$ are the pressure and volume of the adsorbed phase, is much smaller that the two other terms and can be neglected. For the supported layer, $\big|N_{\sigma}~\mu_{\rm v}\big|_{0}^{5\times10^4}=\big|N_{\sigma}\,k_{B}T~ln(P)\big|_{0}^{5\times10^4}=1.21$~J which shows that the corresponding variation of the surface free energy, $-0.21$~J, cannot be neglected. Concerning the membrane, $\big|N_{\sigma}\,k_{B}Tln(P)\big|_{0}^{5\times10^4}=1.08$~J and $\big|\gamma A\big|_{0}^{5\times10^4}=-0.19$~J. We see that the variation of the total free energy, $F_{\sigma,\rm so}$, of the system adsorbate-porous solid in the pressure range $0-5\times10^{4}$~Pa is lower for the membrane, $0.89$~J, than for the supported layer, $1$~J. This can be interpreted as follows: in its relaxed state (the membrane) the porous system has three degrees of freedom and can find a new configuration for the solid atoms which decreases the total free energy ($\sim10\%$) and the adsorbed amount ($10\%$) and increases the condensation pressure with regard to the supported layer.

(2) We now compare the estimated value of the coefficient $k_{sl}$ to other experimental results found in the literature. For adsorption of water in porous glass, \cite{Amberg52,Grosman08-II} the variation of the surface free energy in the region preceding capillary condensation equals $\sim-17$~J/g and the corresponding relative extension, measured by Amberg \textit{et al.}, \cite{Amberg52} is $\Delta l/l\sim1.2\times10^{-3}$ which leads to $k=1.4\times10^{4}$~J/g. The $k$ values estimated in a porous silicon layer from theoretical considerations independently of adsorption data, $k_{sl}=1.8\times10^{4}$~J/g, and in porous glass from the adsorption data and relative extension measurements, $k=1.4\times10^{4}$~J/g, are hence similar. This indicates that for both materials the relative deformations during adsorption should be approximatively in the same ratio as the surface free energy variations. Note that if porous glass is assummed to be submitted to isotropic strain the coefficient $k$ is related to Young's modulus of porous glass $E_{pgl}$ and to Poisson coefficient $\nu_{pgl}$ by the classical relation $k=(3/2)~[E_{pgl}/(1-2\nu_{pgl})]~V_{\rm so}$. Inserting $\nu_{pgl}\simeq0.2$ \cite{Scherer} we find $E_{pgl}=14$~GPa while for porous silicon Ashby relation gives $E_{p}\simeq41$~GPa which is coherent if we consider that bulk silicon is stiffer than bulk glass. \cite{Scherer}

For pentane in porous silicon, \cite{Dolino96} the relative extension, $\delta a/a$, is small, a few $10^{-5}$, before condensation occurs, while in the hysteresis region it equals a few $10^{-4}$. This experimental result is qualitatively similar to that obtained for water in Charcoal \cite{Wiig49}: in both cases the elastic deformations are much more important in the hysteresis region. For water in Charcoal this behavior is clearly related to the shape of the hysteresis: the adsorbed amount before capillary condensation is very small. This should also be the case for pentane in porous silicon. In our experiments, the amount of nitrogen molecules adsorbed before condensation nearly corresponds to half of the total pore volume (see Figure~\ref{fig:layermemb}). This can explain why the value calculated above, $(\delta a/a)_{sl}=3.2\times10^{-4}$, is higher than that found for pentane before condensation.

(3) Let us comment on the relation $k_{m}=\beta~k_{sl}$ and the value $\beta=1.2$ found by Barla \textit{et al.} \cite{Barla}. For supported layers we have seen that $k_{sl}\simeq E_{p}~V_{\rm so}$. For membranes, under the assumption of isotropic deformation, we should have $k_{m}=(3/2)~[E_{p}/(1-2\nu_{p})]~V_{\rm so}$. By inserting $\nu_{p}\simeq0.1$ we find $k_{m}\simeq1.9$~$k_{sl}$. Note that the low value $\nu_{p}=0.1$ corresponds to a solid submitted to high anisotropic strains. For isotropic material $\nu_{p}$ is certainly higher so that the actual value of $\beta$ may be even larger than $1.9$. Similar $\beta$ values have been found for epitaxic layer on (100) substrate for which the lattice mismatch measured in the $\left[100\right]$ direction has the same value in the two other directions. For example, for epitaxic layer Al$_{x}$Ga$_{1-x}$As on (100) GaAs substrate, for a lattice mismatch of $4.86\times10^{-4}$, $\beta =2.2$. \cite{Estop}

The fact that, for porous silicon, the coefficient $\beta$ is close to $1$ indicates that this porous material does not behave elastically neither as a cubic epitaxial layer on silicon nor as an isotropic porous medium. It is rather an orthotropic system in which the properties are the same in the ($Ox$) and ($Oy$) directions parallel to the interface but different in the ($Oz$) direction perpendicular to the interface. The deformations perpendicular to the pore axis (transversal deformations) are much smaller than the deformations parallel to the pore axis (longitudinal deformations).

Let us note that, in ref~20, the Poisson's coefficient value $\nu_{p}\simeq 0.1$ has been deduced from the relation \cite{Hornstra}
\begin{equation}
\bigg(\frac{\Delta a}{a}\bigg)_{sl}=\beta~\bigg(\frac{\Delta a}{a}\bigg)_{m}=\frac{1+\nu_{p}}{1-\nu_{p}}\bigg(\frac{\Delta a}{a}\bigg)_{m}.
\end{equation}
This relation, based on the assumption of a homogeneous material, cannot be applied to p$^{+}$-type porous silicon. Nevertheless, if we consider the high anisotropy of the material, such a small value around $0.1$ seems quite plausible.

\begin{figure}
\begin{center}
\includegraphics[width=8.5 cm]{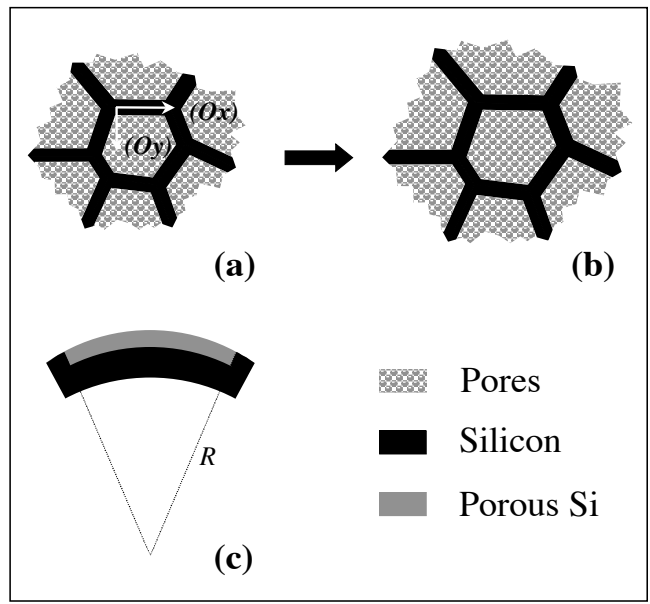}
\end{center}
\caption{Schematic representation of the effects of the relaxation of the stress exerted by the substrate on a pore in a plane perpendicular to the pore axis when detaching a supported porous layer (a) to obtain a membrane (b): the pore perimeter increases, and the wall thickness is unchanged. The lattice mismatch between the membrane and silicon produces an elastic bending of the substrate (c), which becomes convex with a radius of curvature $R$.}
\label{fig:bending}
\end{figure}
We can give a qualitative description of the deformation of the membrane compared to that of the supported layer. Parallel to the pore axis ($Oz$) the extension is probably uniform for both layers. In the plane perpendicular to the pore axis, the interatomic spacing in the supported porous layer is the same as that of the substrate while for the membrane the situation is more complicated. In this plane, for a given pore wall, two directions must be considered (see Figure~\ref{fig:bending}a). Parallel to the pore walls ($Ox$), the surface stress exerted by the chemisorbed molecules (e.g., SiH$_{x}$ groups) increases the interatomic spacing and thus the pore perimeter with regard to those of the supported porous layer. Perpendicular to the pore walls ($Oy$), the lattice strains are different. Indeed, the contraction of the pore walls along ($Oz$), compared to the supported layer, involves an extension of the pore wall along ($Oy$), proportional to Poisson's coefficient. On the other hand, the extension of the pore wall along ($Ox$) involves a contraction along ($Oy$). Considering that the contraction of the pore length and the increase of the pore perimeter have opposite effects on the pore wall thickness and that Poisson's coefficient is small, we can conclude that the wall thickness should be similar for both layers. The effects of the relaxation of the stress exerted by the substrate on a pore are schematically represented in Figures~\ref{fig:bending}a and b. In summary, perpendicular to the pore axis, the elastic deformation in the relaxed state of porous silicon (the membrane) is not uniform. Removing the substrate increases the pore size but lets the pore wall thickness remain unchanged. It is perhaps why the $\beta$ value is close to $1$.

As seen below, the stresses at the interface porous layer-substrate produce an elastic bending of the substrate which becomes convex (see Figure~\ref{fig:bending}c). The supported porous layer behaves like an epitaxial layer over silicon which would have a positive lattice mismatch. We shall come back to this problem in the next section.

Let us conclude this section now.

(1) The supported porous layer and the membrane react differently to the spreading pressure exerted by the adsorbed molecules. The deformation is unilateral (along the pore axis) for the supported layer, while the membrane is also submitted to transversal deformations. To this new configuration for the solid atoms corresponds a lower total free energy which results in a decrease of the adsorbed amount for the membrane compared to the supported layer.

(2) One can explain why the condensation pressure is higher for the membrane than for the supported layer without assuming any effect of a pore end on capillary condensation. Let us assume that the pores of the supported layer fill in the same way as the membrane. As the vapor pressure is increased, the adsorbed film reaches a critical thickness above which it becomes unstable. A liquid bridge is then formed somewhere in the pores, leading to the filling of the pores; however, since, for a given vapor pressure, the adsorbed film thickness is lower for the membrane than for the supported layer, the filling of the pores will occur at higher pressure, as it is observed. The experimental results presented in the next section support this idea according to which the gap between the two condensation branches is only due to the different elastic deformations to which the supported layer and membrane are submitted.

(3) Consider the two evaporation branches in Figure~\ref{fig:layermemb}: they are not superimposed, whereas the two layers are identical and the pore cross section does not vary with depth. As already noted, under the assumption of an inert solid and whatever the evaporation mechanism proposed, the presence of a second end should not change the position of the evaporation branch. The fact that the supported layer and the membrane empty at different pressures is hence due to the different elastic deformations undergone by the two layers. This supports our thermodynamic approach \cite{Grosman08-II} according to which the energy barrier to evaporation is essentially due to the elastic deformation of the porous matrix.

Recently, Bruschi et al. \cite{Bruschi} studied gas adsorption in nanoporous alumina which is charaterized by a regular arrangement of noninterconnected cylindrical pores. For $25$~nm pore diameter, the authors report on a similar result as that shown in Figure~\ref{fig:layermemb}, i.e., the hysteresis loop obtained in pores open at both ends is shifted to a higher relative pressure than that observed for closed-bottom pores. They attribute this shift to a slight widening of the pores when preparing the porous membrane. By considering the present study, their results could be explained in the same way as ours.
\subsection{Stress at the porous layer-compact silicon interface}
\label{subsec:interfstress}
\paragraph{}
The fact that the $100$~$\mu$m thick supported porous layer behaves like a membrane during the adsorption-desorption process (see Figure~\ref{fig:isot100}) suggests that the thick layers are imperfectly supported by the substrate, that is, Si-Si bonds break at the bottom of the pores under the effect of stresses generated at the interface between the porous layer and the substrate, leading to the formation of dislocations. However, as shown below, the layer is not detached from the substrate.

\begin{figure}
\begin{center}
\includegraphics[width=8.5 cm]{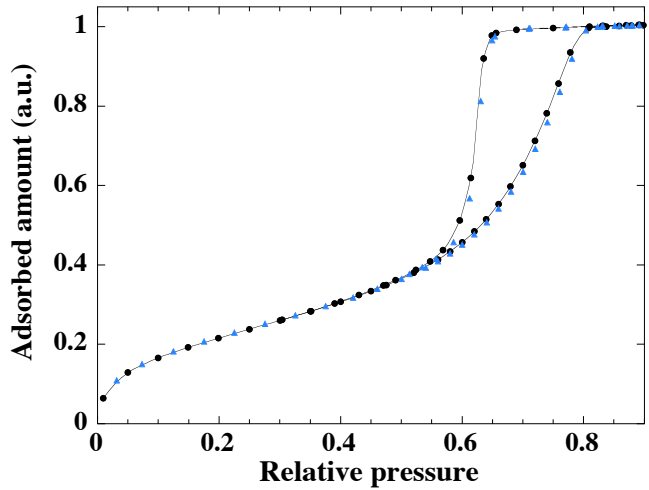}
\end{center}
\caption{Nitrogen adsorption isotherms at $77.4$~K for $100$~$\mu$m thick porous silicon layer (symbol:$\bullet$) and a $100$~$\mu$m thick porous silicon layer chemically etched down to $\sim20$~$\mu$m (symbol:$\blacktriangle$), both supported by a silicon substrate. The lines are guides for the eye.}
\label{fig:abraded}
\end{figure}

First of all, we have verified that the breaking of Si-Si bonds occurs during the formation process and not during the adsorption experiments under the effect of additional stresses or of low temperature (77~K), and so on. We have made a $100$~$\mu$m porous layer and thinned it in a NaOH/H$_{2}$O solution to obtain a $20$~$\mu$m thick layer. Let us recall that dried porous silicon is hydrophobic so that the solution does not enter the pores, which allows us to thin the porous layer without changing the pore size. Figure~\ref{fig:abraded} shows the corresponding isotherm together with that of a $100$~$\mu$m supported porous layer. Once normalized to the same pore volume, the two isotherms are superimposed, which shows that the Si-Si bonds break during the formation process. The formation of thick porous layers (e.g., $100$~$\mu$m) changes irreversibly the porous layer-substrate interface.

For what physical reasons do Si-Si bonds break at the porous silicon-substrate interface when making thick porous layers? The formation of a porous silicon layer in a silicon wafer produces an elastic bending of the sample, which becomes convex. \cite{Barla} As already noted, the system behaves like an epitaxial layer over silicon which would have a positive lattice mismatch, ($\Delta a/a$)$_{m\bot}$, perpendicular to the pore axis. The substrate radius of curvature is given by \cite{Reinhart}
\begin{equation}
R=(d+d_{p})\left[1+\frac{1}{3}\Big(\frac{s}{d}+\frac{s_{p}}{d_{p}}\Big)\Big(\frac{d^{3}}{s}+\frac{{d_{p}}^{3}}{s_{p}}\Big)(d+d_{p})^{-2}\right]\times\left[\frac{\Delta a}{a}\right]_{m\bot}^{-1},
\end{equation}
where $2d=t$ and $2d_{p}=t_{p}$ are the thicknesses of the substrate and of the supported porous layer. $s=(1-\nu)/E$ and $s_{p}=(1-\nu_{p})/E_{p}$ where $E, E_{p}$ and $\nu, \nu_{p}$ are Young's moduli and Poisson's coefficients of silicon and of the porous layer, respectively. The porous layer thickness is measured by weighing as described in ref~15. Remember that the elastic deformation in the relaxed state of porous silicon (the membrane) is not uniform so that $(\Delta a/a)_{m\bot}$ is not really a lattice mismatch as is the case for epitaxial layers, but only a phenomenological parameter: $(\Delta a/a)_{m\bot}$ can be considered as an "effective" lattice mismatch. 

In our case, we have ($t+t_{p}$)$\simeq300$~$\mu$m and $t_{p}\lesssim100$~$\mu$m. In these conditions, as verified below, eq~13 can be replaced with a good approximation by
\begin{eqnarray}
R&=&\frac{1}{6}\;\frac{s_{p}}{s}\frac{1}{(\Delta a/a)_{m\bot}}\frac{t^{2}}{t_{p}}\!\!\!\!\!\nonumber\\
\!\!\!&=&\!\!\frac{1}{6}\;\frac{E}{E_{p}}\;\frac{1-\nu_{p}}{1-\nu}\;\frac{1}{(\Delta a/a)_{m\bot}}\;\frac{t^{2}}{t_{p}}.
\end{eqnarray}
The difference ($\sigma_{2}-\sigma_{1}$) between the surface stresses exerted on the front side of the wafer where the porous layer is formed and on the back side is given from classical elasticity theory by
\begin{equation}
R=\frac{1}{6}\;\frac{E}{(1-\nu)}\;\frac{1}{(\sigma_{2}-\sigma_{1})}\;t^{2}.
\end{equation}
%
\begin{figure}
\begin{center}
\includegraphics[width=8.5 cm]{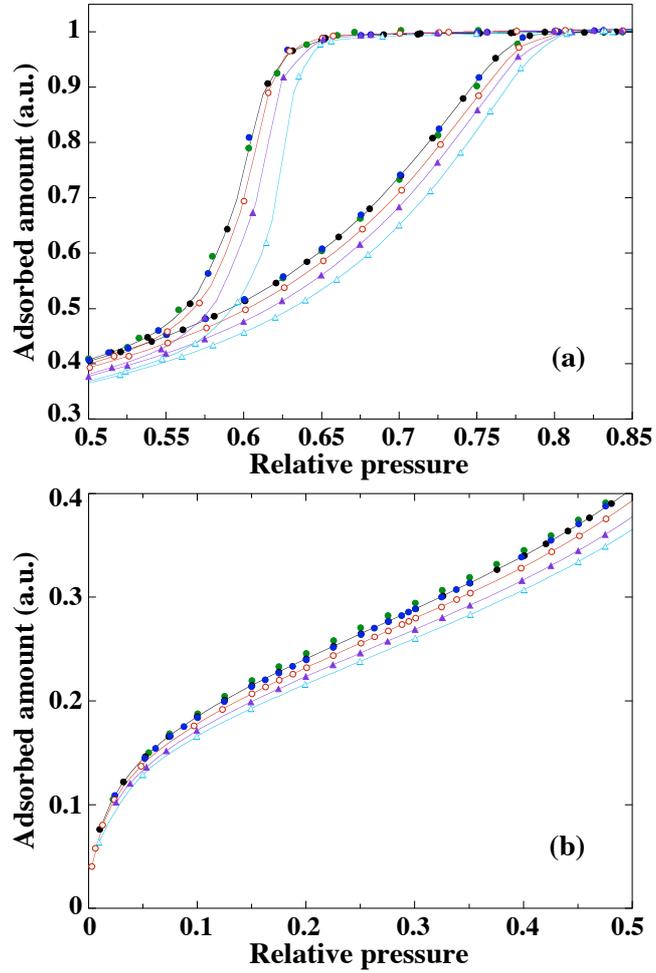}
\end{center}
\caption{Nitrogen adsorption isotherms at $77.4$~K for $10$~$\mu$m, $20$~$\mu$m, $30$~$\mu$m, $40$~$\mu$m, $50$~$\mu$m and $100$~$\mu$m thick supported porous layers in (a) the hysteresis region and (b) the pressure range $[0-0.5]$. The isotherms for layers $10$, $20$, and $30$~$\mu$m (symbols:$\bullet$) thick are superimposed in the whole pressure range. The lines are guides for the eye.}
\label{fig:isothick}
\end{figure}
From eqs~14 and 15, we get 
\begin{equation}
(\sigma_{2}-\sigma_{1})= \frac{E_{p}}{(1-\nu_{p})}\;(\Delta a/a)_{m\bot}\;t_{p}.
\end{equation}
$(\sigma_{2}-\sigma_{1})$ is hence proportional to the thickness of the porous layer. Increasing the porous layer thickness increases the stresses at the porous layer-substrate interface and can cause Si-Si bonds to break. This effect is illustrated in Figure~\ref{fig:isothick}, which shows N$_{2}$ adsorption isotherms for porous layers of increasing thickness from $10$ to $100$~$\mu$m in the hysteresis region (Figure~\ref{fig:isothick}a) and in the low pressure range (Figure~\ref{fig:isothick}b). Up to a thickness of $\simeq30$~$\mu$m, the whole isotherms are superimposed, then they shift toward that corresponding to the membranes, the adsorbed amount in the region of reversible adsorption preceding the capillary condensation decreasing continuously (see Figure~\ref{fig:isothick}b). These results indicate that Si-Si bonds begin to break only when the thickness exceeds a limit between $\simeq30$~$\mu$m and $\simeq40$~$\mu$m. From this limit, as the thickness is increased, more and more Si-Si bonds break so that the stress exerted by the substrate on the porous layer is less and less observable in adsorption experiments. Consequently, the broken Si-Si bonds are not localized only at the porous layer-substrate interface but also in the whole porous layer from a depth $> 30$~$\mu$m.
\begin{figure}
\begin{center}
\includegraphics[width=8.5 cm]{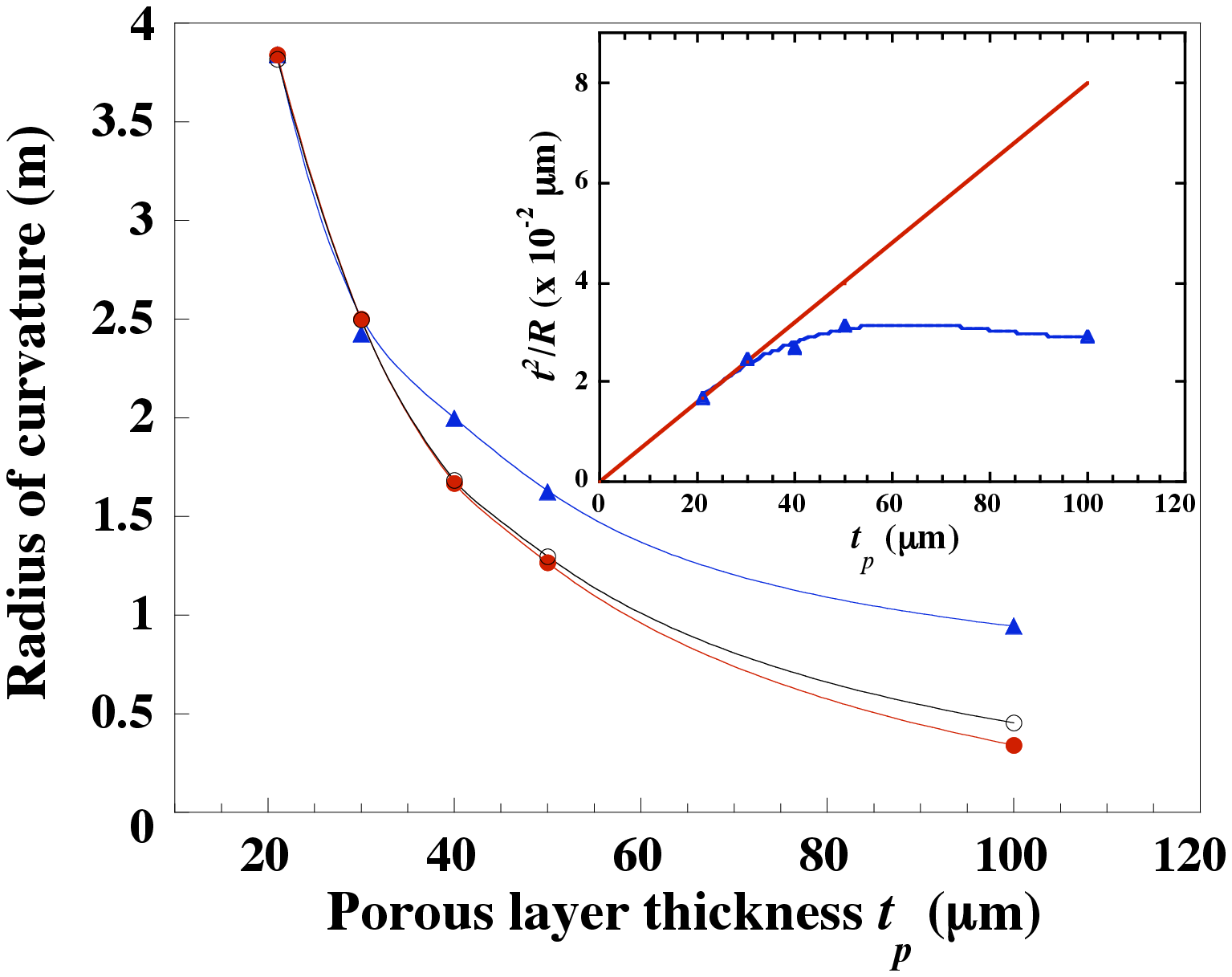}
\end{center}
\caption{Radius of curvature $R$ measured from Newton's rings (symbol:$\blacktriangle$) and calculated from eqs~13 (symbol: $\circ$) and 14 (symbol:$\bullet$) as a function of the porous layer thickness, $t_{p}$. The $R$ values correspond to several measurements with a deviation of $\sim2\%$. The inset shows the experimental values of $t^{2}/R$ as a function of $t_{p}$, where $t$ is the substrate thickness (symbol:$\blacktriangle$). The straight line which passes by the origin and the first two experimental points is calculated from eq~14. The porosity is $50\%$.}
\label{fig:Rcourbure}
\end{figure}

To confirm this explanation, we have determined optically the radius of curvature of the different samples by measuring the diameters of Newton's rings. The results are shown in Figure~\ref{fig:Rcourbure} as a function of the porous layer thickness. We have found that the Newton's rings are circular for thick porous layers ($50$ and $100$~$\mu$m) but elliptical with two orthogonal principal axes for thinner layers. For example, for the $20$~$\mu$m thick layer, we found $R_{1}\sim3.5$~m and $R_{2}\sim4$~m. The $R$ values in Figure~\ref{fig:Rcourbure} correspond to several measurements with a deviation of $\sim2\%$. As expected, the radius of curvature of porous layers decreases with increasing thickness.

We have checked whether these measurements can be fitted by the theoretical predictions. Since Si-Si bonds do not break in the $20$ and $30$~$\mu$m thick porous layers, the curves represented by eqs~13 and 14 must pass by the points corresponding to these two layers. In the case of eq~14, the only parameter which we need, $(s_{p}/s)(\Delta a/a)_{m\bot}^{-1}$, has been deduced from the radius of curvature of one or the other of the two layers. In the case of eq~13, we need to know separately the values of the two parameters, $(s_{p}/s)$ and $(\Delta a/a)_{m\bot}^{-1}$. As we have seen above, porous silicon is an anisotropic material. In eqs~13 and 14, the elastic compliance parameter ($s_{p}$) is related to deformations parallel to the interface. This parameter which can be different from that corresponding to deformations perpendicular to the interface is unknown. However, we have estimated ($s_{p}$) by assuming that $E_{p}=E~(1-P_{or})^{2}$, $\nu_{p}\simeq0.1$ which corresponds to deformations perpendicular to the interface. For bulk silicon, $\nu=0.26$. \cite{Barla} The second parameter, $(\Delta a/a)_{m\bot}$, has been deduced as above from the radius of curvature of the $20$~$\mu$m thick layer. We have found $(\Delta a/a)_{m\bot}=6.5\times10^{-4}$. If our qualitative description represented schematically in Figure~\ref{fig:bending} is good, this value should represent an estimate of the extension mean value of the pore diameter when the stress exerted by the substrate is relaxed.

The two theoretical curves corresponding to eqs~13 and 14 are represented in Figure~\ref{fig:Rcourbure}. As noted above, we see that eq~14 is a good approximation of eq~13. The radius of curvature of layers thicker than $30$~$\mu$m deviates more and more, in relative terms, from the theoretical values, which shows that an increasingly large number of Si-Si bonds break when increasing the thickness from $30$ to $100$~$\mu$m.
\begin{figure}
\begin{center}
\includegraphics[width=8.5 cm]{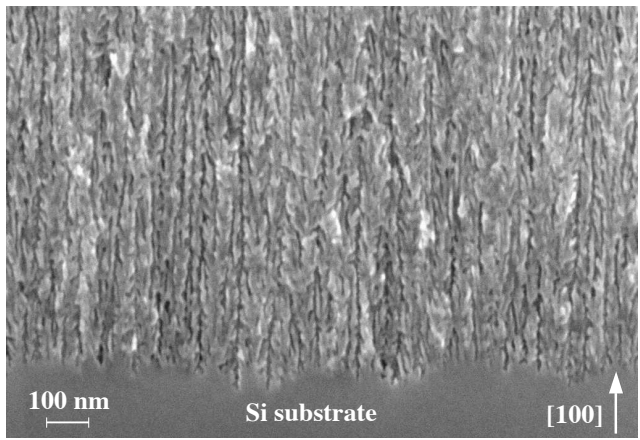}
\end{center}
\caption{Scanning electron microscopy image in cross section of the porous silicon-silicon substrate interface for a porous layer $100$~$\mu$m thick with a porosity of $50\%$. The amplitude of the undulations of the porous layer-substrate interface is about $100$~nm, that is, $10^{-3}$ in relative terms.}
\label{fig:MEB}
\end{figure}

This phenomenon is even more visible in the inset of Figure~\ref{fig:Rcourbure} which shows the relationship $(t^{2}/R)$ versus $t_{p}$. The first two points which correspond to the $20$ and $30$~$\mu$m thick layers are on a straight line which passes by $0$ according to eq~14. Then, once the Si-Si bonds begin to break, the corresponding experimental points deviate strongly from this linearity. The results shown in Figures~\ref{fig:isothick} and \ref{fig:Rcourbure} are quite coherent. The shift of the adsorption isotherms (Figure~\ref{fig:isothick}) appears to have a bearing on the concomitant decreasing of the radius of curvature (Figure~\ref{fig:Rcourbure}).

Whereas the $100$~$\mu$m thick layer behaves like a membrane in adsorption-desorption process, it is not completely detached from the substrate, since the radius of curvature has a finite values ($\sim1$~m). To check whether these breakings are visible, we have performed scanning electron microscopy observations. The image of the porous layer-substrate interface shown in Figure~\ref{fig:MEB} did not reveal any fracture at a scale of $\sim100$~nm.

When the constraint of the substrate is partially relaxed by the breaking of Si-Si bonds, the isotherms coresponding to the $100$~$\mu$m thick supported layers are superimposed to that of the membranes. This indicates that the PSDs for both types of layers are similar, which justifies the assumption we made at the end of Section~\ref{subsec:layerandmemb}. This suggests also that the difference between the two condensation branches observed for the thin layers (see Figure~\ref{fig:layermemb}) is not due to pore end effect but has the same cause as the difference between the adsorbed amounts before condensation. The gap between the two condensation branches is entirely due to the different elastic deformations undergone by the thin supported layer and the membrane.

Before discussing our results, let us comment recent experiments made on porous silicon layers similar to ours. \cite{Naumov} In this work, the adsorption isotherms for $30$~$\mu$m thick supported layer and membrane have been found to be superimposed, which is in contradiction with our results. It is surprising that the authors who refer to our previous work published in 2002, \cite{CoasnePRL} which shows, for the first time, that the isotherms for $20$~$\mu$m thick supported layer and membrane are distinct, did not mention at all the disagreement between the two experimental results. 

How to explain such a disagreement? 

1) The process used to detach the porous layer from the substrate is not described in ref~\cite{Naumov} but we suppose that it is the same as that currently used in the literature and by us, that is, an increase of the current intensity after the pores reach the desired length. We have seen above that this process lets the porous layer unchanged. Thus, in both experiments, the supported layer and membrane are identical.

2) The (100) Si substrates used in both experiments are similar, $2-5$~m$\Omega$.cm in ref~\cite{Naumov} and $2.7-3.3$~m$\Omega$.cm in the present paper, like as the current density ($20$~mA/cm$^{2}$). However, the etching solutions were different. In ref~\cite{Naumov}, the etching solution was HF($48\%$)/EtOH=1:1 while in our case it was HF($40\%$)/EtOH=3:1. The HF and water concentrations were hence different in the two experiments and consequently the porosity and etching rate which depend importantly on these parameters (see below).

(i) Concerning the porosity, according to the calibrations we made, the conditions chosen in ref~\cite{Naumov} should lead to porous layers with a porosity around $60\%$. Thus, the layers studied in ref~\cite{Naumov} do not seem to be very different from ours which is coherent with the fact that the superimposed hysteresis loops for the membrane and supported layer are observed at pressures only slightly higher than represented in Figure~\ref{fig:isot100}.

(ii) The thickness of the porous layers was stated to be $30$~$\mu$m but the authors did not tell us if they measured the thickness by weighing or if they chose the formation time from our calibrations. \cite{Grosman08-I} The etching rate depends importantly on the water and HF concentrations. For a given current density, for instance $50$~mA/cm$^{2}$, the etching rate for HF($40\%$)/EtOH=1:1 is $1.7\mu$m/min whereas for HF($48\%$)/EtOH=1:1 it equals $2.5\mu$m/min, that is $1.5$ times higher. We have shown in this paper that the film thickness is a determinant parameter for breaking of Si-Si bonds like the stiffness and the effective lattice mismatch which depend on the porosity: for the porous layers we studied, the critical thickness from which the Si-Si bonds begin to break equals $30$~$\mu$m. What is the critical thickness for the layers of ref~\cite{Naumov}? In the absence of more informations we cannot say anything more than: if indeed the isotherms for supported porous layer and membrane in ref~$\cite{Naumov}$ are superimposed, the apparent disagreement between this result and ours represented in Figure~\ref{fig:layermemb} should have the same explanation as the apparent disagreement between our Figures~\ref{fig:layermemb} and \ref{fig:isot100}, that is, the partial relaxation of the constraint of the substrate when making thicker layers. 

\section{Discussion}
\label{sec:discussion}
\paragraph{}
We have seen in this paper that the isotherms corresponding to porous layers supported by the substrate and the corresponding membranes are distinct whereas these layers are strictly similar. If we take into consideration the elastic deformation of the porous solid, these isotherms correspond in fact to different experiments.

The porous silicon layer, when attached to the substrate, is submitted, during the adsorption process, to unilateral deformation perpendicular to the substrate, that is, along the pore axis (longitudinal deformations). The problem is equivalent to the classical case of the compression of a block held by lateral walls in such a way that its tranversal dimensions may be regarded as constant. When the porous layer is detached from the substrate to obtain a membrane, deformations perpendicular to the pore axis (transversal deformations) can also occur so that the extension along the pore axis is lower. The longitudinal deformations depend on the "effective" elastic constants, $k_{m}$ for the membrane and $k_{sl}$ for the supported layer, which connect the variation of the surface free energy with the deformations in the [100] direction. These elastic constants are related one to the other by the relation $k_{m}=\beta k_{sl}\simeq1.2k_{sl}$. The fact that $\beta\simeq1.2$ is close to $1$, or, in an equivalent way, that Poisson's coefficient has a low value ($\nu_{p}\simeq0.1$), reflects the high anisotropy of the porous system. Indeed, $\beta =1$ and $\nu_{p}=0$ would correspond to a porous system whose transversal dimensions are unaffected by the deformation along the pore axis. On the other hand, for isotropic porous material such as porous glass, \cite{Scherer} taking $\nu_{p}\simeq0.15-0.2$ we obtain $\beta\gtrsim 2$.

The relaxation of the constraint of the substrate allows the porous system to jump from a state corresponding to one-dimensional deformations to another state corresponding to tridimensional deformations. Although the transversal deformations are low compared to the longitudinal deformations since the two type of layers have very similar "effective" elastic constants along the [100] direction, the change of state corresponds to a significant decrease of the total free energy ($\sim10\%$). This decrease results in a decrease of the adsorbed amount before condensation ($10\%$) accompanied by a shift of the condensation pressure toward higher pressure, typically $0.7\rightarrow0.73$.

Let us now consider a hypothetical porous system which could jump from an initial state corresponding to unilateral deformation as it is the case for the supported layer, to a final state corresponding to uniform tridimensional deformation as it is the case for porous glass or silica aerogel ($\beta\gtrsim 2$). Our results suggest that, for the isotropic system, the decrease of the adsorbed amount before condensation should be much higher than $10\%$ and the condensation pressure much higher than $0.73$.

In the light of this interpretation, let us analyze some results concerning other porous systems such as MCM-41 or SBA-15 silica, which are periodic arrangements of cylindrical pores. For porous silicon and the SBA-15 material that we have studied, \cite{Grosman05} the relative condensation pressures are similar, $p\simeq0.7$, while the mean pore size is $\simeq13$~nm for porous silicon and $\simeq8.5$~nm for SBA-15 silica. In a previous paper, \cite{CoasnePRL} we had noted that the mean condensation pressure for porous silicon is lower than that expected from the classical Kelvin equation (inert adsorbent), roughly $0.7P_{0}$ instead of $0.9P_{0}$. Moreover, calculations based on density functional theory \cite{Malanoski-I, Malanoski-II} failed to reproduce such low condensation pressures. The condensation pressure for porous silicon seems hence "abnormally" low compared to that of SBA-15 silica having similar pore size and to theoretical values found under the assumption of an inert adsorbent. Note that to quantitatively reproduce the actual size for large pores in MCM-41 ($6-7$~nm) it is necessary to apply the Kelvin equation for a hemispherical meniscus to the capillary condensation branch. \cite{Kruk} This means that, as for porous silicon, the condensation pressure is lower than that expected from the classical Kelvin equation, i.e., inert adsorbent and cylindrical meniscus.

What do we know about the elastic deformation of MCM-41 and SBA-15? For such porous materials, the only parameter which can be measured is the distance between the pore lattice planes, that is, the transversal deformations, so that we cannot compare the longitudinal and transversal deformations. Concerning the transversal deformations, there are still few results \cite{Zickler, Gunther} but they all suggest isotropic deformation. Moreover, Poisson's coefficient values for these materials \cite{Zickler, Fan} are between $0.17$ and $0.24$, values close to that of homogeneous materials and importantly higher than that of porous silicon ($\nu_{p}\simeq0.1$). We conclude that these materials react to the surface pressure exerted by the adsorbate in a similar way as isotropic materials. According to the above interpretation, this can partly explain the gap between the condensation pressures for porous silicon and SBA-15.

Other parameters influence the condensation pressure such as the adsorbate-porous solid interactions or the pore morphology. We can note, for example, that the pores of porous silicon have a polygonal shape more or less regular whereas those of SBA-15 seem to be quite cylindrical. The PSD centered on $\simeq13$~nm in porous silicon layers under study corresponds to cylindrical pores having the same cross section surface area as the polygonal pores. If, as it is probably the case, the condensation in a given pore occurs when a liquid bridge is formed somewhere in the pore, the important parameter is the minimum distance between two walls. To estimate this distance, we have determined the PSD of the circles inscribed in the polygones: it is centered on $11.5$~nm. Thus, porous silicon is perhaps not equivalent to an assembly of cylindrical pores with a diameter mean value equal to $13$~nm but rather to $11.5$~nm.

The theoritical calculations performed under the assumption of an inert adsorbent fail to reproduce the "abnormally" low condensation pressures for porous silicon, while the disagreement for materials submitted to isotropic strains (SBA-15) seems to be less important. This does not mean that the influence of elastic deformation is less important for material submitted to isotropic strains than for material submitted to anisotropic strains. Porous silicon (high anisotropic strains) and porous glass (isotropic strains) have very similar qualitative hysteretic behavior with similar boundary hysteresis loops and subloops inside the main loop. The difference between the total free energies for the condensation and evaporation branches for both systems is essentially due to the difference between the surface free energies, that is, the elastic deformations, which suggests that the height of the barrier to evaporation is essentially due to the elastic deformation. \cite{Grosman08-II} For a given pore size, the position of the hysteresis loop on the pressure axis can depend on whether the porous system is submitted to anisotropic or isotropic strains, but the hysteretic behaviors are qualitatively similar.

An alternative explanation based on the assumption of inert adsorbent has been proposed in the literature to account for the triangular shape of hysteresis loop in porous silicon, \cite{Naumov, Wallacher, Joel} while in SBA-15 or MCM-41 the shape is rather rectangular with steep and parallel condensation and evaporation branches. The hysteretic behavior of porous silicon would be due to the presence of high disorder in the pores, fluctuation of pore diameter, \cite{Naumov} chemical heterogeneities, \cite{Joel} and so on. One of the consequences of the hypothetical disorder introduced in each pore is that condensation is nucleated via the formation of liquid bridges so that pore ends do not play any role on the condensation process. Thus, the simulated isotherms for pores closed at one end (the supported layer) and for pores open at both ends (the membrane) are superimposed, \cite{Joel, Naumov} which is in contradiction with our experimental results. The disorder cannot explain why the experimental isotherms for the supported layer and membrane are distinct.

On the other hand, we point out that our results suggest strongly that capillary condensation is nucleated via the formation of liquid bridges and not at the bottom of the pores. Indeed, we have seen that, in the region preceding capillary condensation, the normalized adsorbed amount at a given pressure decreases continuously with increasing layer thickness to reach the value corresponding to that of the membrane. We have explained this phenomenon by the partial relaxation of the constraint exerted by the substrate on the porous layer. Simultaneously, we observe that the condensation branch shifts toward higher pressures to finally superimpose itself on that of the membrane. The simultaneity of the two phenomena suggests that they are correlated. The question is whether the shift of the condensation branch observed when the thinnest supported layers are detached from the substrate (Figure~\ref{fig:layermemb}) could be partly due to pore end effect as modeled by Cohan. \cite{Cohan38} In the $100$~$\mu$m thick supported layer, the major part of Si-Si bonds are not broken, since the curvature radius is small ($1$~m). Thus, the pores, or at least the major part of them, are closed at the porous layer-substrate interface. The pore size over pore length ratio is very small in the $20$~$\mu$m thick layer, of the order of $5/10000$, so that, if the formation of liquid bridges does not occur in the first $20$~$\mu$m, there is no reason that it occurs at higher depth, since the average pore section does not vary with depth. Thus, the simple fact that the condensation branches for the $100$~$\mu$m thick supported layer and membranes are superimposed suggests strongly that there is no pore end effect on the condensation process.

The calculations of Naumov and co-workers, \cite{Naumov} reproduced a hysteresis loop with a triangular shape similar to that experimentally observed in porous silicon. The geometry they considered consists of a slit pore divided in segments, with each segment having a diameter in the range $4-8$ lattice constant. For the porous silicon layers studied in the present paper, it would mean that each pore is composed of segments having diameter in the range $9-17$~nm. In other words, the pore size distribution we determined from TEM images in plane view would represent the variation of the diameter of each pore. This should result in large undulations of the walls of each pores in opposite directions. These undulations should let the pore wall thickness remain unchanged, $5-6$~nm, as well as the axes of the pores parallel one to the other. Indeed, we have shown by proton energy loss fluctuations measurements that the average direction of the pores is identical to within $0.1^{¡}$ of the $\left[100\right]$ crystal axis. \cite{Amsel} We have never observed such undulations.

The calculations presented in ref~\cite{Naumov} are based on an unfounded hypothesis, and the fact that they reproduce the shape of the experimental sorption isotherms does not prove at all that the matter looks like the model, especially if these calculations are based on an unjustified assumption: an inert porous solid.

To explain the presence of both a triangular hysteresis loop and an interaction mechanism in porous silicon where the pores are not interconnected, \cite {CoasnePRL} we follow a different way. We realized that the assumption of an inert adsorbent in adsorption process is unjustified. \cite {Grosman08-II} Let us consider, for example, the variation of the total free energy at the top of the plateau between the condensation and desorption branches. We can see that the variation of the surface free energy, deduced from elastic deformation measurements, for porous glass and porous silicon, is of the same order of magnitude as the other terms of the total free energy and cannot thus be neglected. To neglect the elastic deformation, that is, the variation of the surface free energy, it would lead to a very small variation of the total free energy with regards to the chemical and mechanical energies of the fluid.
\section{Conclusion}
\label{sec:concl}
\paragraph{}
The linear deformation of a porous material during the adsorption-desorption process is proportional to the surface free energy, the proportionality factor depending on the elastic constants of the porous solid. \cite{Amberg52, Bangham30, Dash76} This relation is the result of two equilibrium conditions, one thermodynamic and the other mechanic. \cite{Grosman08-II} This means that a stress external to the porous layer can change the adsorbed amount and hence the condensation pressure. It is what we show in the present paper. The external stress is here exerted by the substrate on the porous layer. With porous silicon one can make identical porous layers either supported by or detached from the substrate. Whereas the two types of layers are identical, the isotherms are distinct. During gas adsorption in the supported porous layers, the planes perpendicular to the substrate are forced to have the same interatomic spacing as that of the substrate so that the porous layer is submitted to unilateral deformation, while in its relaxed state (the membrane) the porous system has three degrees of freedom and can find a new configuration for the solid atoms which significantly decreases the total free energy ($\sim10\%$) and the adsorbed amount ($\sim10\%$) and increases the condensation pressure ($0.7\rightarrow0.73$) compared to that of the supported layer.

The stress exerted by the substrate can be partly relaxed by making thicker porous layers. The relaxed lattice parameter of a porous silicon membrane is greater than that of silicon, \cite{Barla} so that the supported porous layer behaves like an epitaxial layer over silicon, which would have a positive lattice mismatch: the formation of a porous layer in a Si wafer produces an elastic bending of the wafer which then becomes convex. As the porous layer thickness is increased, the stress at the porous layer-substrate interface increases and the radius of curvature decreases although it remains higher than expected by the theory. This is due to the breaking of Si-Si bonds (dislocations) caused by the stress at the interface. The stress exerted by the substrate is thus partly relaxed by the formation of dislocations, which results in the shift of the isotherms toward that of the membrane when the porous layer thickness is increased from $30$ to $100$~$\mu$m.

The longitudinal deformations (along the pore axis) depend on the "effective" elastic constants, $k_{m}$ for the membrane and $k_{sl}$ for the supported layer, which connect the variation of the surface free energy with the deformations in the [100] direction. These elastic constants are related one to the other by the relation $k_{m}=\beta k_{sl}\simeq1.2k_{sl}$. The large disagreement between this experimental value ($\beta\simeq1.2$) and that found in porous material submitted to isotropic deformation ($\beta\gtrsim 2$) indicates clearly that perpendicular to the pore axis the elastic deformation in the relaxed state of porous silicon (the membrane) is much lower than longitudinal deformations. Although the transverval deformations are weak, the relaxation of the substrate results in a significant decrease of the total free energy and the adsorbed amount quoted above.

This suggests that, for a porous system which would be subjected to an uniform tridimensional deformation ($\beta\gtrsim 2$), the decrease of the adsorbed amount and the shift of the condensation pressure toward higher value should be much higher than those observed for the porous silicon membrane. This can partly explain why the condensation pressure for porous silicon layers is low compared to that of other ordered porous materials such as MCM-41 and SBA-15 silica having similar pore size, since these latter materials seem to react to the surface pressure exerted by the adsorbate in a similar way as isotropic materials.

This does not mean that the influence of elastic deformation is less important for material submitted to isotropic strains than for material submitted to anisotropic strains. The hysteretic behaviors for porous silicon (anisotropic strains) and porous glass (isotropic strains) are very similar, and the height of the barrier to evaporation is essentially due to the elastic deformation for both systems. The position of the hysteresis loop on the pressure axis depends on whether the porous system is submitted to anisotropic or isotropic strains, but the hysteretic behavior is qualitatively similar.

The experimental results presented in this paper show the significant influence of stresses external to a porous system on the adsorption process. This supports recent Monte Carlo simulations \cite{Gunther} which show that the elastic strains affect the fluid's phase behavior. This supports also our thermodynamic approach and the interaction mechanism proposed, \cite{Grosman08-II} in which the external mechanical stress is imposed on a given pore by the neighbors.

\textbf{Acknowledgments.} We acknowledge F. Bernardot (INSP) for performing optical measurements of the radius of curvature of porous layers and for fruitful discussions and S.~Borensztajn (LISE, UPR 15, CNRS, France) for the SEM observations. 


\end{document}